\documentstyle[floats,aps,aipbook,psfig]{revtex}

\def\etal{et\ al.}
\def\g{{$\gamma$}}

\begin{document}

\title{A Constraint on the Distance Scale to\\
    Cosmological Gamma--Ray Bursts}

\author{Jean M. Quashnock\thanks{{\it Compton} GRO Fellow}}

\address{Department of Astronomy and Astrophysics \\
        University of Chicago, Chicago, Illinois 60637}

\maketitle
\begin{abstract}

If \g--ray bursts have a cosmological origin,
the sources are expected to trace the large--scale structure
of luminous matter in the universe.
I use a new likelihood method that compares the
counts--in--cells distribution of \g--ray bursts 
in the BATSE 3B catalog with that
expected from the known large--scale structure of the universe,
in order to place a constraint on the distance scale to cosmological bursts.
I find, at the 95\% confidence level, that the comoving distance to the
``edge'' of the burst distribution 
is greater than $630~h^{-1}$~Mpc ($z > 0.25$),
and that the nearest burst is farther than $40~h^{-1}$~Mpc.
The median distance to the nearest burst is $170~h^{-1}$~Mpc,
implying that the total energy released in \g--rays during a burst event 
is of order $3\times 10^{51}~h^{-2}$ erg.
None of the bursts that have been observed by BATSE are in nearby galaxies,
nor is a signature from the Coma cluster or the ``Great Wall'' likely to be
seen in the data at present.

\end{abstract}

\section*{Introduction}

The origin of \g--ray bursts is still unknown
and is currently the subject of a ``great debate''
in the astronomical community.
Do the bursts have a galactic origin \cite{L95}\ or
are they cosmological \cite{Pac95} ?
And what is their distance scale?

In this paper, I do not attempt to answer the first question,
but rather, I show that {\it if} one assumes that \g--ray bursts 
are cosmological in origin,
one can begin to answer the second question and place a
constraint on the distance scale to the bursts.
This is because cosmological bursts
are expected to trace the large--scale structure
of luminous matter in the universe \cite{LQ93} .
The constraint comes from
comparing the {\it expected} clustering pattern of bursts on the sky
--- which will depend on their distance scale because of projection effects ---
with that {\it actually observed}. 
The observed angular distribution is in fact quite isotropic \cite{Brig95} ;
hence, only a lower limit to the distance scale can be placed
because a sufficiently large distance will always lead to 
a sufficiently isotropic distribution on the sky.

Here I use a powerful new likelihood method \cite{Q96} ,
which I had previously developed to analyze repeating of
\g--ray bursts in the BATSE 1B and 2B catalogs \cite{Q95a} ,
to compare the observed counts--in--cells distribution
in the new BATSE 3B catalog \cite{Meeg95}\ with that expected
for bursts at cosmological distances.

Here I will assume for simplicity that $\Omega_0=1$ and $\Lambda=0$,
and that the large--scale structure clustering 
pattern is constant in comoving coordinates.
The results are in fact insensitive to these assumptions because of the
small redshifts that are involved.
I follow the usual convention and take $h$ to be the Hubble constant
in units of 100 km~s$^{-1}$~Mpc$^{-1}$.

\section*{Likelihood Method}

Let $N_{\rm cell}$ be a large number of circular cells,
each centered on a random position on the sky.
Each cell is of fixed solid angle size $\Omega=2\pi(1-\cos\theta_{\rm rad})$,
where $\theta_{\rm rad}$ is the angular radius of the cell.
I set the number of cells to be such that any part of the sky is covered,
on average, by one cell; hence, $N_{\rm cell}=4\pi/\Omega$.
Let $C_N$ to be the number of these cells having $N$ \g--ray bursts in them,
out of the $N_{\rm tot}=1122$ in the BATSE 3B catalog,
where $N=0,1,2,...$  I then define the observed counts--in--cells distribution,
$P_N\equiv C_N/N_{\rm cell}$, 
as the probability that a randomly chosen cell of size $\Omega$
has $N$ bursts in it.
The counts--in--cells distribution contains information about clustering
of \g--ray bursts on scales comparable 
to the angular size $\theta_{\rm rad}$ of the cell.  

I now define $Q_N$ to be the counts--in--cells distribution that is expected
if \g--ray bursts are cosmological in origin 
and trace the large--scale structure of luminous
matter in the universe.
This expected distribution depends on only one unknown parameter,
the effective distance $D$ to \g--ray bursts (which I define below),
because the angular clustering pattern of bursts on the sky will depend
by projection on this distance.
 
The {\it likelihood} ${\cal L}$ measures how likely it is that 
the observed counts--in--cells distribution $P_N$
is drawn from the expected distribution $Q_N$.
Since $Q_N$ depends on the unknown effective distance $D$ to \g--ray bursts,
the likelihood is really a measure of how likely a given value of $D$ is.
I find that \cite{Q96} :
\begin{equation}
\log{\cal L} = N_{\rm cell} \sum_N P_N \log Q_N + {\rm constant}\; .
\end{equation}

Now the cumulative $C_{\rm max}/C_{\rm min}$ distribution
of \g--ray bursts seen by BATSE begins to roll over from a $- 3/2$ power--law
for bursts fainter than $C_{\rm max}/C_{\rm min} \sim 10$.
Since this is many times above threshold,
it suggests that BATSE sees most of the source distribution
and that this distribution is not spatially homogeneous \cite{Meeg92} .
I define $D$ as the comoving distance beyond which 
the source density drops appreciably.
It is not the distance to the very dimmest burst in the BATSE catalog, 
but rather the typical distance to most of the dim bursts in the sample; 
thus, $D$ is the {\it effective distance}
to the ``edge'' of the source distribution in the BATSE catalog. 

I take the power spectrum which characterizes the large--scale
clustering of \g--ray burst sources 
to be the same as that determined from a 
redshift survey of radio galaxies \cite{PN91} .
This power spectrum is characteristic of moderately rich environments,
and is intermediate between that of ordinary galaxies and clusters.
Because the exact bias factor relating the clustering of \g--ray burst sources 
to that of luminous matter is unknown,
such an intermediate ansatz is reasonable.
In any case, the resultant distance limit depends only weakly
on the bias factor (roughly as the square root).

Knowledge of the power spectrum permits a calculation of the expected
angular clustering pattern, the expected counts--in--cells distribution $Q_N$,
and finally the likelihood $\cal L$ [from equation (1)],
all as a function of the effective distance $D$ to \g--ray bursts \cite{Q96}.
I have included the smearing due to finite positional errors
on the clustering on small scales \cite{HLB91} .
Indeed, each burst in the BATSE catalog 
is assigned a positional uncertainty $\theta_{\rm err}$
corresponding to a $68\%$ confidence that the true burst position
is within an angle $\theta_{\rm err}$
to the position listed in the catalog.

I have chosen the cell size
$\theta_{\rm rad}$ in order to maximize the sensitivity of detection,
or signal--to--noise,
given the strength of the signal expected.
For a sample of 1122 bursts (the total number of bursts in the
BATSE 3B catalog) with positional smearing of $\theta_{\rm err}=3.8^\circ$ 
(the median value in the 3B catalog),
the signal--to--noise is maximized when cells of $\theta_{\rm rad}=5^\circ$
are used \cite{Q96} .

\section*{Results}

Figure 1 shows the likelihood
of the BATSE 3B catalog data as a function
of the effective comoving distance $D$,
calculated using cells of size $\theta_{\rm rad}=5^\circ$.
The likelihood is normalized to that expected for an isotropic
distribution on the sky.
At large values of $D$ (the maximum value allowed 
is $D=R_{\rm H}=6000~h^{-1}$ Mpc,
the size of the horizon in a closed universe),
the likelihood goes to unity, because
by projection a sufficiently large distance will always lead to an isotropic
distribution on the sky. Note also that there is no value of $D$ for which
the likelihood is greater than 1; 
thus, the maximum likelihood value for $D$ is $R_{\rm H}$
and the 3B data are consistent with isotropy.

The solid line in Fig. 1
shows the likelihood for a positional smearing of $\theta_{\rm err}=3.8^\circ$,
corresponding to the median value in the 3B catalog.
To illustrate the dependence of these results on
positional errors, I also show (dashed line) the results 
for a larger positional smearing\footnote{Graziani \& Lamb \cite{GL95}\
compare the 3B positions with those
from the IPN network, 
and conclude that the systematic errors are larger than the $1.6^\circ$
value quoted in the 3B catalog. 
Their best--fit model gives a median positional error of $6.6^\circ$.}  
of $\theta_{\rm err}=6.6^\circ$
(with cells of size $\theta_{\rm rad}=9^\circ$ 
to maximize signal--to--noise).

Small values of the effective comoving distance to
\g--ray bursts are unlikely, according to Fig. 1:
I find, at the 95\%
confidence level, that for the 3B median positional error
of $\theta_{\rm err}=3.8^\circ$,
$D$ must be greater than $630~h^{-1}$ Mpc,
corresponding to a redshift $z > 0.25$.
If the positional errors are larger than quoted 
and are better characterized
by $\theta_{\rm err}=6.6^\circ$, these results are only slightly
weakened:
At the 95\% confidence level,
$D$ must be greater than $500~h^{-1}$ Mpc,
corresponding to a redshift $z > 0.19$.

These limits are not sensitive to earlier
assumptions on cosmology and clustering
evolution since these only become important at
higher redshifts.
They are also conservative limits in that a constant median
value for the positional errors was used, rather than the entire
distribution of errors.
This is because the bright bursts, which ostensibly are
nearer to us, are more clustered and are responsible for
the bulk of the expected signal, but in fact have smaller
errors than the median value. The faint bursts, which
are far away, are hardly clustered to begin with
(even before smearing), but have errors larger than the median value.
Hence the expected clustering pattern has been smeared more
by using a constant median value (this permits a simpler calculation)
than by smearing using
the entire distribution of errors.
So the counts--in--cells statistic has been weakened somewhat and
thus the quoted lower limits are in fact conservative.

\begin{figure}
\centerline{\psfig{file=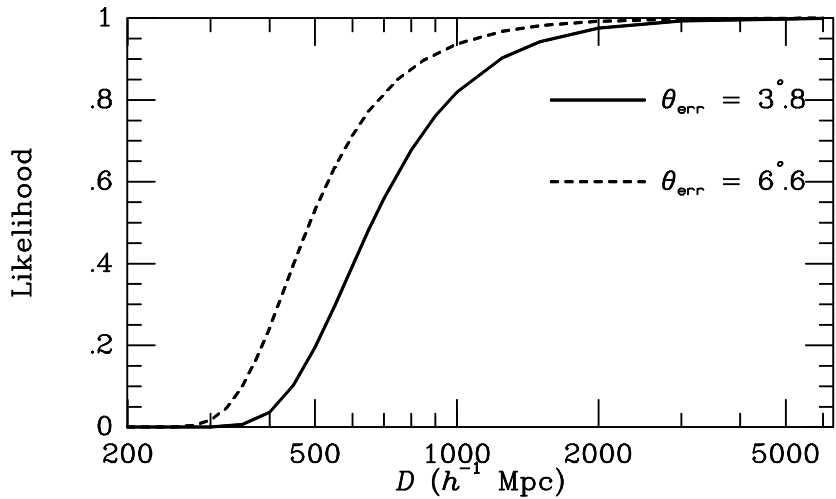,width=11cm,angle=0}}
\bigskip
\caption{The likelihood of the BATSE 3B catalog data as a function
of the effective comoving distance $D$ to \g--ray bursts,
shown with a smearing of $\theta_{\rm err}=3.8^\circ$ and $6.6^\circ$.}
\end{figure}

\section*{Conclusions}

If \g--ray bursts are cosmological and trace the large--scale
structure of luminous matter in the universe,
and their  positional errors are as quoted in the 3B catalog,
then the lack of any angular clustering in the data
implies that the observed distance to the ``edge'' of the  burst distribution
must be farther than $630~h^{-1}$ Mpc.
Since there are 1122 bursts in the catalog, 
an effective limit on the {\it nearest} burst to us can be
placed by convoluting the likelihood as a function of $D$ (Fig. 1) with
the nearest neighbor distribution of 1122 bursts inside a sphere of
radius $D$. I find that the nearest burst
must be farther than $40~h^{-1}$ Mpc at the 95\% confidence level,
and farther than $10~h^{-1}$ Mpc at the 99.9\% level.
At this level of confidence, then, none of the bursts that have been
observed by BATSE are in nearby galaxies.
A signature from the Coma cluster or the ``Great Wall'' 
($\sim 70~h^{-1}$ Mpc) is not likely to be
seen in the data at present, since only a few bursts
could have originated from these distances.

The median distance to the nearest burst is $170~h^{-1}$ Mpc.
Since the brightest burst in the 3B catalog
has a fluence of $7.8\times 10^{-4}~{\rm erg~cm}^{-2}$ in \g--rays,
this implies that the total energy released in \g--rays during a burst event 
is of order $3\times 10^{51}~h^{-2}$ erg.

As the number of observed \g--ray bursts keeps increasing,
the distance limit will improve.
In fact, with 3000 burst
locations, the clustering of bursts might just be detectable \cite{LQ93}\ 
and would provide compelling evidence for a cosmological origin.
If it is not detected, the redshift to the ``edge'' of the burst
distribution  would be put at $z\sim 1$ or beyond.

\acknowledgments

I would like to acknowledge useful discussions with Carlo Graziani,
Don Lamb, Cole Miller and Bob Nichol.
This research was supported in part by NASA through the {\it Compton}
Fellowship Program --- grant NAG 5-2660, grant NAG 5-2868,
and contract NASW-4690.

\end{document}